\documentclass[aps,amsmath,amssymb,reprint, superscriptaddress]{revtex4-2}

\setcitestyle{super}

\usepackage{xcolor}
\usepackage{graphicx}% Include figure files
\usepackage{dcolumn}% Align table columns on decimal point
\usepackage{bm}% bold math
\usepackage[utf8]{inputenc}
\usepackage[T1]{fontenc}

\usepackage{txfonts}

\usepackage{siunitx}
\usepackage{etoolbox}
\usepackage{array}
\newcolumntype{P}[1]{>{\centering\arraybackslash}p{#1}}

\begin{document}

\title[]{Structural features and nonlinear rheology of self-assembled networks of cross-linked semiflexible polymers}
\author{Saamiya Syed}
\affiliation{ 
College of Technology, University of Houston, Houston, TX 77204, USA
}%
\affiliation{ 
Center for Theoretical Biological Physics, Rice University, Houston, TX 77005, USA
}%
\author{Fred C. MacKintosh}%
\affiliation{ 
Center for Theoretical Biological Physics, Rice University, Houston, TX 77005, USA
}%
\affiliation{ 
Department of Chemical and Biomolecular Engineering, Rice University, Houston, TX 77005, USA
}%
\affiliation{ 
Department of Chemistry, Rice University, Houston, TX 77005, USA
}%
\affiliation{ 
Department of Physics \& Astronomy, Rice University, Houston, TX 77005, USA
}%
\author{Jordan L. Shivers}
\email[]{jshivers@uchicago.edu}
\affiliation{ 
Center for Theoretical Biological Physics, Rice University, Houston, TX 77005, USA
}%
\affiliation{ 
Department of Chemical and Biomolecular Engineering, Rice University, Houston, TX 77005, USA
}%
% \date{\today}

\begin{abstract}
\textbf{Abstract:} Disordered networks of semiflexible filaments are common support structures in biology. Familiar examples include fibrous matrices in blood clots, bacterial biofilms, and essential components of cells and tissues of plants, animals, and fungi. Despite the ubiquity of these networks in biomaterials, we have only a limited understanding of the relationship between their structural features and highly strain-sensitive mechanical properties. In this work, we perform simulations of three-dimensional networks produced by the irreversible formation of crosslinks between linker-decorated semiflexible filaments. We characterize the structure of networks formed by a simple diffusion-dependent assembly process and measure their associated steady-state rheological features at finite temperature over a range of applied prestrains that encompass the strain-stiffening transition. We quantify the dependence of network connectivity on crosslinker availability and detail the associated connectivity dependence of both linear elasticity and nonlinear strain stiffening behavior, drawing comparisons with prior experimental measurements of the crosslinker concentration-dependent elasticity of actin gels.
\end{abstract}

\maketitle
\thispagestyle{plain}

\section{Introduction
}

The formation of living things involves the energy-intensive assembly of complex structures from scarce resources.  Survival requires that these structures remain robust and functional under significant and often repetitive applied stresses and strains. Quasi-one-dimensional or filamentous structures address these challenges efficiently by supporting significant tensile stresses with minimal material cost.\cite{burla_mechanical_2019} Biological polymers and fibers are generally also semiflexible,\cite{janmey_mechanical_1991} meaning that they resist modes of deformation that induce bending, such as applied compression. In examples spanning a wide range of length scales, including information-storing DNA and RNA, actin and intermediate filaments in the cell cytoskeleton, and extracellular collagen and elastin fibers in tissues, these mechanical features are essential for biological function. 

The cytoskeleton and extracellular matrix are examples of disordered networks, a common class of higher-order structures in living materials. Building on the qualities of their underlying semi-flexible filaments, these networks act as responsive elastic scaffolds that resist extreme deformation while leaving ample space for the transport and storage of functional components, such as interstitial fluids and cells. Unlike conventional elastic solids, their mechanical properties are scale-dependent\cite{tyznik_length_2019,proestaki_modulus_2019} and sensitive to changes in applied stress or strain,\cite{onck_alternative_2005,picu_poissons_2018,ban_strong_2019}  to which they respond with dramatic stiffening, alignment, and changes in local filament density, enabling essential biological phenomena such as long-range force transmission by cells\cite{notbohm_microbuckling_2015,wang_long-range_2015,han_cell_2018,alisafaei_long-range_2021,grill_directed_2021} and muscle fiber contraction.\cite{huxley_mechanism_1969,gunst_first_2003} Recent work has suggested that the non-linear viscoelastic properties of these and related fibrous networks are governed by an underlying mechanical phase transition\cite{sharma_strain-controlled_2016,jansen_role_2018} associated with the onset of stretching-dominated rigidity under applied shear or extensile strains. Within this framework, elastic properties are predicted to exhibit a power-law dependence on applied strain in the vicinity of a critical strain, and nonaffine (inhomogeneous) rearrangements that become increasingly large near the critical strain are expected to drive a significant slowing of stress relaxation as the transition is approached. \cite{shivers_nonaffinity_2022} This slowing occurs due to the physical coupling of the rearranging elastic network to a viscous background (the solvent) and is closely related to the well-known divergence of the  viscosity of dense particulate suspensions near the onset of jamming.\cite{andreotti_shear_2012,lerner_unified_2012} The magnitude of the critical strain, at which these effects are the most pronounced, depends sensitively on key features of the underlying network architecture.\cite{wyart_elasticity_2008, broedersz_filament-length-controlled_2012}

Extensive simulation-based studies have explored the nonlinear rheological properties of disordered networks of crosslinked stiff or semiflexible polymers,\cite{onck_alternative_2005,picu_mechanics_2011,carrillo_nonlinear_2013,freedman_versatile_2017} in some cases with realistic three-dimensional geometries produced either by physical assembly processes\cite{huisman_three-dimensional_2007,kim_computational_2009,dobrynin_universality_2011,muller_rheology_2014,muller_resolution_2015,zagar_two_2015,debenedictis_structure_2020,grill_directed_2021} or artificial generation procedures.\cite{huisman_monte_2008,lindstrom_biopolymer_2010,huisman_internal_2011,huisman_frequency-dependent_2010,lindstrom_finite-strain_2013,amuasi_nonlinear_2015} However, efforts to specifically connect network structure with strain-controlled critical behavior have typically focused on simplified random spring networks.\cite{wyart_elasticity_2008,vermeulen_geometry_2017,jansen_role_2018,rens_micromechanical_2018,merkel_minimal-length_2019,shivers_normal_2019,shivers_nonlinear_2020,burla_connectivity_2020,burla_stress_2019,dussi_athermal_2020,tauber_stretchy_2022} In spring networks, the critical strain coincident with the onset of stretching-dominated mechanics can be tuned by changing the average connectivity $z$, defined as the average number of bonds joined at each network junction.\cite{wyart_elasticity_2008} For a network of crosslinked filaments, $z$ is controlled by the typical number of crosslinks formed per filament, approaching an upper limit of $z\to 4$ at high crosslinking density.\cite{broedersz_filament-length-controlled_2012} In biopolymer gels, small changes in the concentration of available crosslinkers can drive dramatic changes in rheological properties,\cite{broedersz_modeling_2014} including changes in the linear elastic modulus \cite{piechocka_multi-scale_2016} and shifts in the critical strain corresponding to the onset of stretching-dominated mechanics.\cite{gardel_elastic_2004}  These changes are naively consistent with a tendency of the connectivity to increase with crosslinker concentration. However, the quantitative relationships between crosslinker concentration and connectivity, and the structural characteristics of assembled networks more generally, remain poorly understood. Improving our knowledge of how the concentration-dependent microscopic structural details of self-assembled networks translate into strain-dependent macroscopic rheological properties is essential to understand the forces at play in important biological processes, such as wound healing and cancer metastasis, and to effectively design biomimetic synthetic materials.\cite{romera_strain_2017}

\begin{figure}[htb!]
\centering
\includegraphics[width=1\columnwidth]{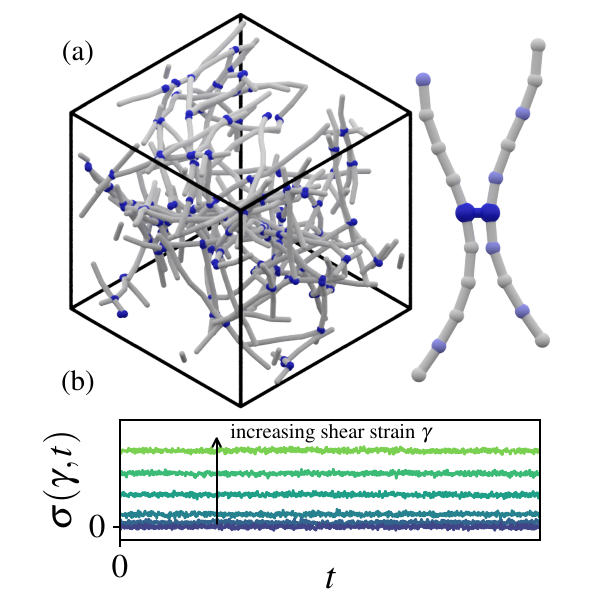}
\caption{ \label{fig_1} Model system and rheological approach. (a) Coarse-grained semiflexible filaments are decorated with randomly assigned sticky sites (light blue spheres in the image on the right) with coverage fraction $p$. When two sticky sites meet, they are connected by a permanent crosslink (royal blue dumbbell). Solutions of these filaments diffusively self-assemble into percolating disordered networks with macroscopic elasticity. (b) After the assembly process is stopped, the network topology remains fixed and the strain-dependent rheological properties are determined from the fluctuating shear stress $\sigma(\gamma, t)$ measured under a simple macroscopic shear strain $\gamma$.}
\end{figure}

In this study, we consider a system composed of coarse-grained semiflexible filaments that diffusively self-assemble into a system-spanning network through the formation of permanent inter-filament crosslinks. We begin with randomly positioned free filaments with a specified coverage fraction $p$ of "sticky" or linker-decorated sites. The linker coverage fraction serves as a proxy for the ratio of crosslinker and filament concentrations in a real system. Allowing diffusive motion to proceed, we add permanent crosslinks (short elastic bonds, depicted in blue in Fig. \ref{fig_1}a) between sticky sites whose pairwise distance decreases beneath a designated crosslink formation distance.  After the rate of diffusion-driven formation of new crosslinks becomes sufficiently slow (see Section S4 in Supporting Information), we stop the assembly process and permanently fix the topology of the network. That is, any crosslinks that have already formed remain permanent, while any sticky sites that have not formed crosslinks are permanently inactivated. We then analyze the structure of the fixed network, measuring the dependence of various structural features on the linker coverage fraction.  An example network is shown in Fig. \ref{fig_1}a, and a movie depicting the assembly process is provided in Supporting Information.

After fixing the network topology, we transition to the rheology stage, in which we observe the behavior of the network at steady state under constant simple shear strain $\gamma$. We obtain time series measurements of the thermally fluctuating shear stress $\sigma(\gamma, t)$ in the mechanically equilibrated state, as shown schematically in Fig. \ref{fig_1}b. Repeating these measurements over a range of strains for each set of input parameters, we calculate relevant elastic quantities such as the differential shear modulus $K_\mathrm{eq}$  and the critical strain $\gamma_c$, at which the macroscopic system transitions between mechanical regimes dominated by bending and stretching. Appropriate physical parameters are chosen to enable comparison with previous experimental measurements of the elasticity of irreversibly crosslinked networks of F-actin,\cite{gardel_elastic_2004} an essential component of the cytoskeleton of eukaryotic cells. Using the same time series measurements, we then characterize the dynamics of stress relaxation via time correlations in the stress fluctuations. Building upon recent work,\cite{shivers_nonaffinity_2022} we demonstrate that the excess differential viscosity, a measure of energy dissipation reflected in a system's finite-temperature stress fluctuations, is directly proportional to the corresponding quasistatic, athermal differential nonaffinity, a measure of the rearrangement induced by a small strain perturbation in the quasistatic, athermal limit. Since in disordered networks the quasistatic nonaffinity is highly strain-dependent and reaches a maximum at the critical strain, analogous behavior is expected in the excess differential viscosity and the slowest viscoelastic relaxation time. Our simulations confirm this expectation, providing crucial insight into potentially measurable effects of nonaffine fluctuations, which have generally proven challenging to experimentally quantify.

\section{Model definition and network assembly}

We imagine a system that begins as a solution of free semiflexible filaments covered to some extent with bound linker proteins that are capable of dimerizing to form permanent elastic crosslinks. In experiments, the linker coverage could be controlled by varying the relative concentrations of the crosslinking protein and filament monomer, as in Ref. \citenum{gardel_elastic_2004}. If the filament concentration and linker coverage in such a solution are sufficiently high, the formation of crosslinks between diffusively migrating filaments eventually produces a macroscopic network.

To capture this behavior, we turn to a simplified computational model. We specify the number of filaments $N_f$, filament length $\ell_f$, and filament length density $\rho$, which together determine the size of the periodic simulation box, $L=(N_f \ell_f/\rho)^{1/3}$. To ensure that the (initially straight) filaments do not span the entire system, we require $L>\ell_f$, or equivalently $\ell_f < \sqrt{N_f/\rho}$. We also specify the number of evenly spaced nodes per filament, $n$, which determines the filament bond rest length $\ell_0 = \ell_f/(n-1)$.  The total number of nodes in the system is then $N = N_f n$. We then designate a fraction $p$ of the nodes, randomly chosen, as sticky or capable of forming a crosslink bond with another node of the same type. Crosslinks are permanent bonds with a rest length $\ell_{0,c\ell}$ shorter than filament bonds. All bonds, including crosslinks, are treated as harmonic springs with stretching stiffness $\mu$, and harmonic bending interactions with stiffness $\kappa = k_B T \ell_p$ act between adjacent filament bonds. Here, $\ell_p$ denotes the persistence length of the filament. In terms of the $3N$-dimensional vector of node positions $\mathbf{x}$, we write the elastic potential energy of this system as
\begin{equation}
    U(\mathbf{x})=\frac{\mu}{2}\sum_{ij}\frac{\left(\ell_{ij}-\ell_{ij,0}\right)^2}{\ell_{ij,0}}+\frac{\kappa}{2}\sum_{ijk}\frac{\left(\theta_{ijk}-\theta_{ijk,0}\right)^2}{\ell_{ijk,0}}
\end{equation}
in which $\ell_{ij}=\left|\mathbf{x}_j-\mathbf{x}_i\right|$, $\theta_{ijk}$ is the angle between the bonds $ij$ and $jk$, $\ell_{ijk}=(\ell_{ij}+\ell_{jk})/2$, and the subscript $0$ denotes the rest values. The first sum is taken over all bonds $ij$, including crosslink bonds, and the second over all pairs of connected bonds $ij$ and $jk$ along the backbone of each filament. Because we consider cross-linked networks with very low filament volume fractions, we deem it acceptable to ignore steric interactions between filaments. Neglecting inertia, as is appropriate for the timescales studied here, the system obeys the overdamped Langevin equation,\cite{allen_computer_2017,frenkel_understanding_2002} such that the forces acting on all nodes satisfy
\begin{equation} \label{eqn:forces}
    \mathbf{F}_N+\mathbf{F}_D+\mathbf{F}_B=\mathbf{0},
\end{equation}
in which the terms on the left represent the network forces, drag forces, and Brownian forces, respectively. The force due to the network is
\begin{equation}
    \mathbf{F}_N=-\frac{\partial U(\mathbf{x})}{\partial \mathbf{x}}.
\end{equation}
The nodes are subjected to a Stokes drag force
\begin{equation}
    \mathbf{F}_D=-\zeta\frac{\partial \mathbf{x}}{\partial t}
\end{equation}
with drag coefficient $\zeta=6\pi\eta_s a$, in which $\eta_s$ is the solvent viscosity and $a$ is an effective node radius of half the filament bond length, $a=\ell_0/2$. Finally, the Brownian force is
\begin{equation}
    \mathbf{F}_B=\sqrt{2\zeta k_B T}\dot{\mathbf{w}},
\end{equation}
in which each component of $\dot{\mathbf{w}}(t)$ is a Gaussian random variable with zero mean and unit variance.\cite{allen_computer_2017} Equation \ref{eqn:forces} can be rewritten as
\begin{equation}\label{eqn:eom}
    \frac{\partial \mathbf{x}}{\partial t}=-\frac{1}{\zeta}\frac{\partial U(\mathbf{x})}{\partial \mathbf{x}} + \sqrt{\frac{2k_B T}{\zeta}}\dot{\mathbf{w}}(t).
\end{equation}

During each timestep in the assembly stage, we check whether the distance separating any pair of sticky nodes has decreased below a specified crosslink formation distance $r_c$, in which case we connect the two with a crosslink bond. Each sticky node can form a maximum of one crosslink, and connections between crosslinks and filaments are treated as freely hinging. In other words, we do not include a bending potential between adjacent crosslink and filament bonds. We stop the assembly stage after a total time $\tau_\mathrm{a}$ has elapsed. We find that $\tau_\mathrm{a} = 6 \times 10^7 \Delta t$, corresponding to approximately one minute in real units, is long enough for the rate of crosslink formation to become negligible (see Section S4 in the Supporting Information for further discussion).  Note that crosslink formation does not take place after the assembly stage; during the rheology stage, described in the next section, the network topology remains fixed.

We note that the assembled networks are inevitably in a prestressed state, as the finite-temperature assembly process involves the formation of new constraints (crosslinks) between filaments that are in fluctuating states of local bending, stretching, or compression. However, as we will see in the next section, the assembled networks clearly remain well within the bending-dominated linear elastic regime, indicating that the effects of this thermally-induced prestress on network elasticity are insignificant.

After assembly, we analyze the structural features of each network. To determine the connectivity $z$, or the average number of connections at each network junction, we consider a reduced version of the simulated network. Each pair of crosslinked nodes in the original network corresponds to a single node in the reduced network, and each filament section between two crosslinked nodes in the original network corresponds to an edge (see sketch in Fig. S1 and further details in Supporting Information, Section S1). Dangling ends, or filament sections in the original network connected to only one crosslink, are therefore neglected. This is a reasonable choice, as dangling ends do not contribute to the elastic response of the network at zero frequency. We then calculate the connectivity from the number of edges $n_\mathrm{edges}$ and nodes $n_\mathrm{nodes}$ present in the reduced network structure as $z=2n_\mathrm{edges}/n_\mathrm{nodes}$.  

We construct systems with filament length $\ell_f = 9 \;\si{\micro\meter}$ and filament persistence length $\ell_p = 17\; \si{\micro\meter}$, chosen to approximate F-actin, with filament length per volume $\rho = 2.6\; \si{\micro\meter}^{-2}$ (for F-actin, this corresponds to a concentration of $c_A = 
1.6\;\si{\micro M}$). Additional parameters are specified in Table S1 (see Supporting Information). Unless otherwise stated, the measurements reported throughout this work are averaged over three randomly generated network samples, and error bars represent $\pm 1$ standard deviation. All simulations are performed using the open source molecular dynamics simulation tool LAMMPS.\cite{thompson_lammps_2022}

\begin{figure}[htb!]
\centering
\includegraphics[width=1\columnwidth]{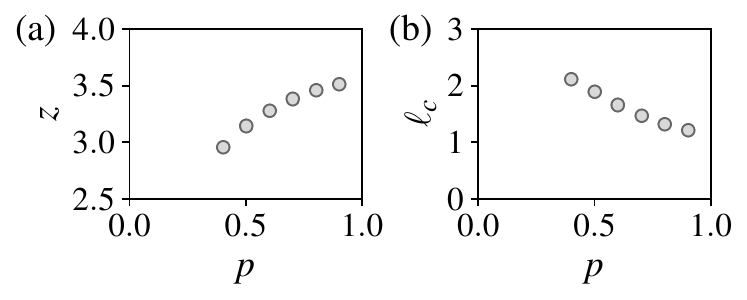}
\caption{ \label{fig_2} Linker coverage-dependence of key structural characteristics of assembled networks. As the linker coverage fraction $p$ increases, we observe (a) an increase in the average connectivity $z$ and (b) a decrease in the average contour length between crosslinks $\ell_c$.}
\end{figure}

We first consider the effects of adjusting the crosslinker coverage fraction $p$ on the assembled network structure. Varying $p$ from $0.4$ to $0.9$, we find that the connectivity of the fully assembled network structures ranges from $z\in[2.8,3.6]$, increasing monotonically with $p$ (see Fig. \ref{fig_2}a). These values are similar to those measured for collagen and fibrin networks in previous experimental work.\cite{lindstrom_biopolymer_2010,beroz_physical_2017,jansen_role_2018,xia_anomalous_2021} Next, we determine the average inter-crosslink contour length $\ell_c$, referring to the average distance between consecutive crosslinks on each filament. We calculate $\ell_c$ as the average length of the filament sections in the original network corresponding to the edges of the reduced network. As shown in Fig. \ref{fig_2}b, we find that $\ell_c$ decreases monotonically with $p$.

\section{Nonlinear rheology}

After the assembly stage is halted, we proceed to the rheology stage, in which the strain-dependent steady-state viscoelastic properties of the system are measured via the time-dependent shear stress as the system fluctuates about the mechanically equilibrated state under a fixed shear strain. We impose a constant macroscopic simple shear strain $\gamma$ using Lees-Edwards periodic boundary conditions \cite{lees_computer_1972} and, using the conjugate gradient method, initially obtain the energy-minimizing configuration of the network, corresponding to the mechanically equilibrated state at $T = 0$. Then, we evolve the system according to Eq. \ref{eqn:eom} at temperature $T>0$, specified in Table S1 of the Supporting Information, for a total run time $\tau_\mathrm{tot} = 3 \times 10^7 \Delta t$, discarding the first half of the trajectory to avoid initialization effects. For a given configuration of the system, we compute the instantaneous virial stress tensor,
\begin{equation}
    \sigma_{\alpha \beta}=\frac{1}{2V}\sum_{i>j}{f_{ij\alpha} r_{ij\beta}}
\end{equation}in which $\mathbf{r}_{ij}=\mathbf{x}_j-\mathbf{x}_i$ and $\mathbf{f}_{ij}$ is the force acting on node $i$ due to its interaction with node $j$. Because we will focus on macroscopic simple shear oriented along the $x$-axis with gradient direction $z$, we now define $\sigma\equiv\sigma_{xz}$ to simplify notation. Once we have obtained a time series of the shear stress $\sigma(\gamma, t)$ at strain $\gamma$, we calculate the time-averaged shear stress $\langle \sigma(\gamma, t)\rangle_t $. After repeating this procedure over many strains in the interval $\gamma\in[0,1]$, we compute the strain-dependent differential shear modulus,
\begin{equation}
    K_\mathrm{eq}(\gamma)=\frac{\partial \langle \sigma(\gamma, t)\rangle_t}{\partial \gamma},
\end{equation} which measures the apparent stiffness of the sample under macroscopic strain $\gamma$ in response to an infinitesimal additional shear strain step. For sufficiently small strains, this yields the corresponding linear shear modulus, $G_0=\lim_{\gamma\to0}K_\mathrm{eq}(\gamma)$. 

\begin{figure}[htb!]
\centering
\includegraphics[width=1\columnwidth]{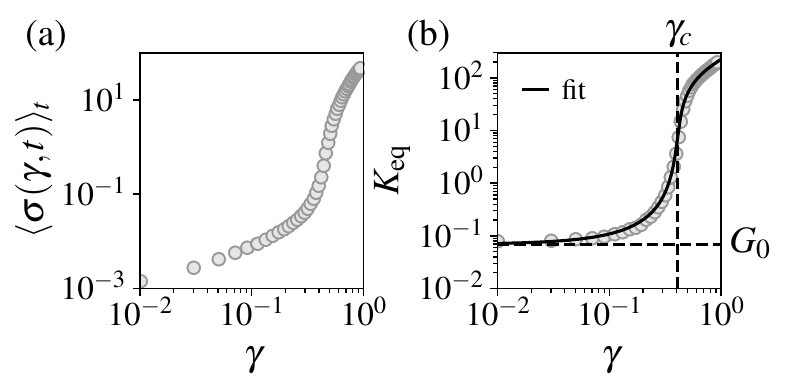}
\caption{ \label{fig_3} (a) Time-averaged shear stress $\langle \sigma(\gamma, t)\rangle_t$ as a function of applied shear strain $\gamma$, for a single network sample with linker coverage fraction $p=0.9$. (b) Differential shear modulus $K_\mathrm{eq}=\partial \langle \sigma(\gamma, t)\rangle_t/\partial \gamma$. The linear shear modulus $G_0$ and the critical strain $\gamma_c$ are indicated by dashed lines, and the solid line is a fit to the equation of state from Ref. \citenum{sharma_strain-controlled_2016}, described in Supporting Information, Section S3. }
\end{figure}

In Figs. \ref{fig_3}a and b, we plot the mean stress $\langle\sigma(\gamma,t)\rangle_t$ and the differential shear modulus $K_\mathrm{eq}$ as a function of strain for a single network sample with the parameters specified in Table S1 (see Supporting Information) and linker coverage fraction $p=0.9$. We determine the critical strain $\gamma_c$, which indicates the transition between the bending-dominated and stretching-dominated mechanical regimes, as the inflection point of the $\log{K_\mathrm{eq}}$ vs. $\log{\gamma}$ curve. To avoid issues associated with differentiating noisy data, we can alternatively find $G_0$ and $\gamma_c$ by fitting the entire $K_\mathrm{eq}$ vs. $\gamma$ curve to an Ising-like equation of state discussed in prior work,\cite{sharma_strain-controlled_2016} as we describe in further detail in Supporting Information, Section S3. Such a fit is shown in Fig. \ref{fig_3}b. For the data presented here, both methods effectively produce equivalent values of $G_0$ and $\gamma_c$.

\begin{figure}[htb!]
\centering
\includegraphics[width=1\columnwidth]{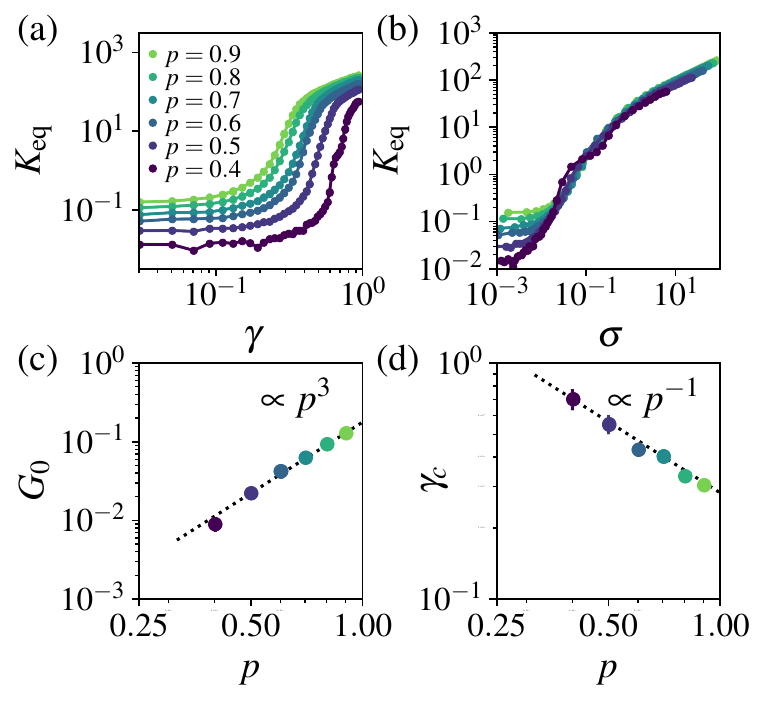}
\caption{ \label{fig_4} Effects of linker coverage on features of the linear and nonlinear stiffness in shear. a) Differential shear modulus $K_\mathrm{eq}$ for networks with varying linker coverage fraction $p$ as a function of strain and (b) as a function of shear stress. Increasing the linker coverage fraction $p$ drives (c) an increase in the linear modulus $G_0=\lim_{\gamma\to 0} K_\mathrm{0}$, with $G_0\propto p^3$, and (d) a decrease in the critical strain $\gamma_c$ as $\gamma_c\propto p^{-1}$. }
\end{figure}

In Fig. \ref{fig_4}, we report the strain dependence of the differential shear modulus for networks with varying linker coverage fraction $p$. As $p$ increases, the critical strain evidently decreases (stiffening occurs at lower applied strains) and the linear shear modulus increases. These observations qualitatively agree with the observed crosslinker concentration dependence of both the linear modulus $G_0$ and the rupture strain $\gamma_\mathrm{max}$ (which we expect to be proportional to $\gamma_c$) of F-actin gels reported in Ref. \citenum{gardel_elastic_2004}, and with the connectivity dependence of the linear modulus and critical strain observed in spring network simulations.\cite{wyart_elasticity_2008} The same differential shear modulus data are plotted in Fig. \ref{fig_4}b as a function of stress. The linear modulus $G_0$ and critical strain $\gamma_c$  values extracted from these curves are plotted as functions of the linker coverage fraction $p$ in Fig. \ref{fig_4}c and d, respectively. We find that each exhibits a power-law dependence on $p$, with $G_0\propto p^3$ and $\gamma_c\propto p^{-1}$. In Ref. \citenum{gardel_elastic_2004}, in which the analogous quantity $R$ (the ratio between the concentrations of actin and the crosslinking protein scruin) is varied, the authors report $G_0\propto R^2$ and $\gamma_c\propto R^{-0.6}$. These observations may be consistent with ours, provided that the linker coverage fraction in our system maps to the experimental crosslinker concentration ratio in Ref. \citenum{gardel_elastic_2004} as $p\propto R^{x}$ with $x\sim 0.6$. Separately, the authors of Ref. \citenum{tharmann_viscoelasticity_2007} reported that, for networks of actin filaments crosslinked by heavy meromyosin, $G_0\propto R^{1.2}$ and $\gamma_c\propto R^{-0.4}$, consistent with our observed dependence of both quantities on $p$ if $x\sim 0.4$. However, meaningfully relating $p$ and $R$ will require further investigation. Here, for example, no two sticky sites on the same filament can reside closer than a distance $\ell_0$ ($1\; \si{\micro\meter}$ with our parameters) from each other. This is obviously not the case in real F-actin networks.

\begin{figure}[htb!]
\centering
\includegraphics[width=1\columnwidth]{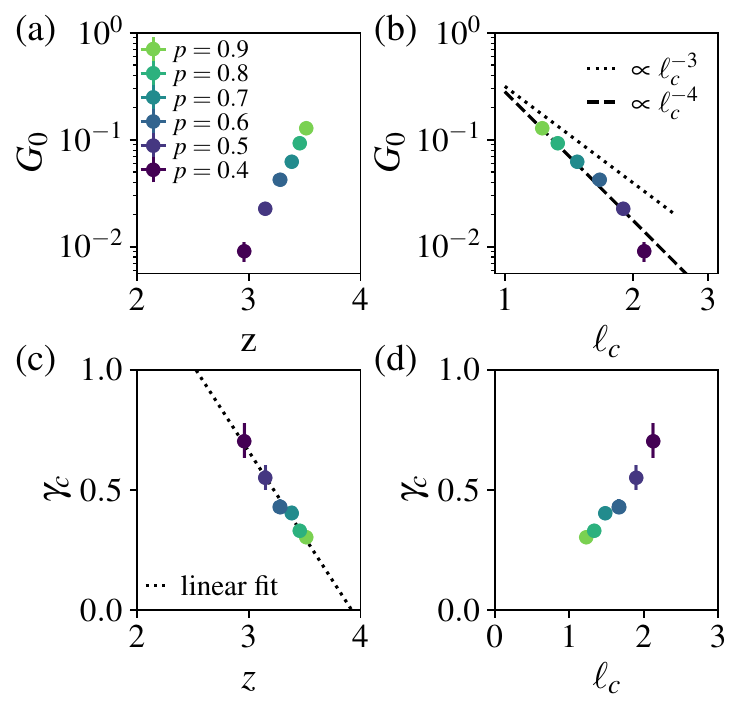}
\caption{ \label{fig_5}  The linear shear modulus $G_0$ (a) increases dramatically with connectivity, shown here for systems with varying linker coverage fraction $p$, and (b) decreases with increasing average inter-crosslink contour length $\ell_c$. For small $\ell_c$ (high $p$), we find $G_0 \propto \ell_c^{-y}$, with $y\sim 4$. Deviation from this scaling appears at higher values of $\ell_c$. (c) The critical strain $\gamma_c$, on the contrary, decreases approximately linearly with $z$. A linear fit suggests $\gamma_c\to0$ as $z\to z^*\sim 4$, roughly consistent with the upper bound for $z$ in the high crosslink density limit.\cite{broedersz_filament-length-controlled_2012} (d) The critical strain increases approximately linearly with increasing average inter-crosslink contour length, $\gamma_c\propto\ell_c$. } 
\end{figure}

In Fig. \ref{fig_5}a and b, we plot the same extracted linear shear modulus measurements as functions of the $p$-dependent structural quantities discussed in the previous section, the average network connectivity $z$ and average inter-crosslink contour length $\ell_c$. We observe an apparent power law dependence of the linear shear modulus $G_0$ on the inter-crosslink contour length $\ell_c$, $G_0 \propto \ell_c^{-y}$, with decay exponent $y\sim 4$. This appears to agree with the expected scaling of $G_0$ with $\ell_c$ for an athermal network of stiff filaments. \cite{huisman_three-dimensional_2007,satcher_theoretical_1996,gibson_cellular_1997} However, it is important to note that our measurements cover less than a decade in $\ell_c$, so any agreement may be coincidental. We thus cannot rule out the $G_0 \propto \ell_c^{-3}$ scaling expected for a thermal semiflexible gel,\cite{mackintosh_elasticity_1995} although for our range of parameters, the typical inter-crosslink contour lengths are presumably too small for bending modes to be properly resolved. Reliably measuring the scaling of $G_0$ with $\ell_c$ in these systems will require further investigation in simulations with reduced coarse graining, i.e., smaller $\ell_0/\ell_f$, over a greater range of $\ell_c$.  In Fig. \ref{fig_5}c, we plot the measured critical strain $\gamma_c$ as a function of the connectivity. We find that, for the range of parameters considered here, $\gamma_c$ appears to exhibit a linear dependence on $z$  with a best-fit intercept near $z^*\sim 4$, in apparent agreement with the upper limit of $z$ for crosslinked networks with high crosslinking density.\cite{broedersz_filament-length-controlled_2012}  It is crucial to note that the dependence of the critical strain on connectivity is known to be sensitive to certain details of the underlying network structure; for example, for short-filament networks with structures derived from 3D jammed sphere packings, the critical strain goes to zero precisely at the isostatic point, $z=6$, and shows a distinctly nonlinear dependence on $z$ far from the isostatic point.\cite{shivers_nonaffinity_2022}  In our networks, we expect the $z$-intercept to depend on the filament length $\ell_f$ and the bending rigidity $\kappa$, which both necessarily influence the structure of the assembled network. While we similarly observe that the critical strain shows an apparent linear dependence on the average inter-crosslink contour length $\ell_c$ (see Fig. \ref{fig_5}d), this too warrants further investigation.

Further information is contained in the fluctuations of the instantaneous stress about its average value,\cite{wittmer_shear_2013}
\begin{equation}
    \delta\sigma(\gamma,t)=\sigma(\gamma,t)-\langle\sigma(\gamma,t)\rangle_t,
\end{equation}
from which we compute the stress fluctuation autocorrelation function,
\begin{equation}
    C(\gamma,\tau)=\frac{V}{k_B T}\langle\delta\sigma(\gamma,t)\delta\sigma(\gamma,t+\tau)\rangle_t,
\end{equation}
which reveals useful details about the time dependence of energy dissipation.\cite{wittmer_fluctuation-dissipation_2015,wittmer_shear-stress_2015}

\begin{figure}[htb!]
\centering
\includegraphics[width=1\columnwidth]{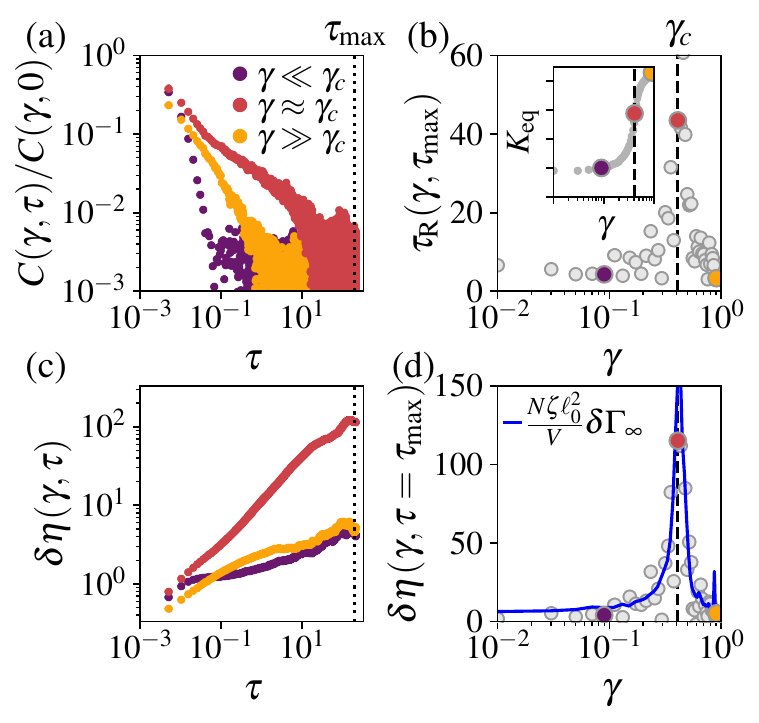}
\caption{ \label{fig_6} The relaxation of stress fluctuations slows dramatically near the critical strain as rearrangements become increasingly nonaffine. (a) The shear stress fluctuation autocorrelation function $C(\gamma, \tau)$ measured at representative strains below (purple), near (red), and above (orange) the critical strain $\gamma_c$ (see the corresponding $K_\mathrm{eq}$ data in the inset of panel (b)) reveals this slowing, which is captured quantitatively by (b) a peak in the apparent slowest relaxation time $\tau_\mathrm{R}$ (Eq. \ref{eqn:relaxationtime}). These data correspond to the same sample as in Fig. \ref{fig_3} (see inset), and the labeled strains are indicated by the same colors in all panels. (c) At long lag times, the time-dependent apparent excess viscosity $\delta\eta(\gamma, \tau)$ calculated with Eq. \ref{eqn:visc} grows much larger near $\gamma_c$ than elsewhere.  This behavior is especially clear when (d) the apparent excess zero-shear viscosity $\delta\eta(\tau=\tau_\mathrm{max})$ is plotted as a function of $\gamma$, revealing a peak at the $p$- (and thus $z$-) dependent critical strain. These observations are supported by independent measurements of the strain-dependent quasistatic, athermal differential nonaffinity $\delta\Gamma_\infty$, which is quantitatively related to $\delta\eta$ by Eq. \ref{eqn:nonaff_viscosity} (solid blue line). }
\end{figure}

\begin{figure}[htb!]
\centering
\includegraphics[width=1\columnwidth]{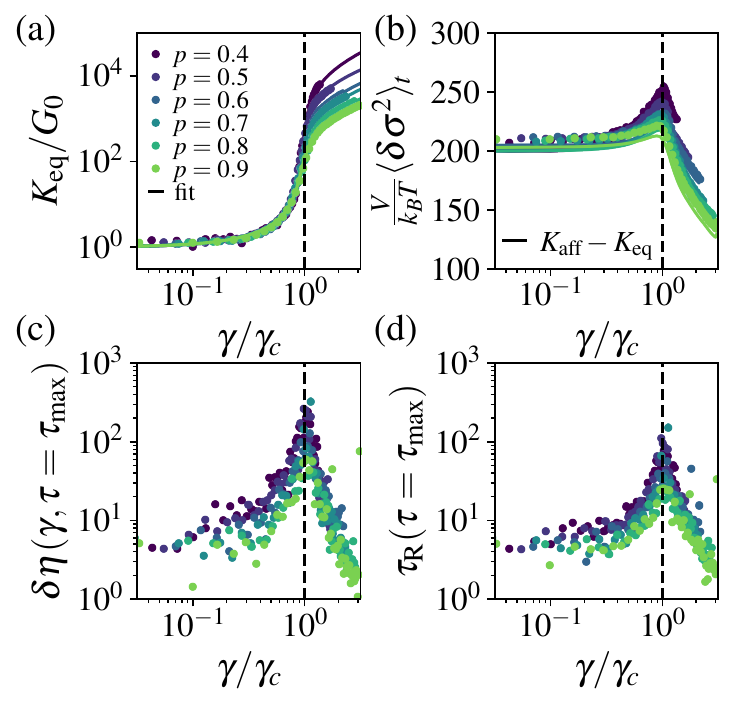}
\caption{ \label{fig_7} Shared features of stiffening and slowing down at the structure-dependent critical strain. (a) For networks with varying crosslinker coverage fraction $p$, the stiffening regime of the normalized differential shear modulus $K/G_0$ collapses onto a single curve when plotted vs. $\gamma/\gamma_c$. Lines correspond to fits to the equation of state described in Ref. \citenum{sharma_strain-controlled_2016} (see Supporting Information, Section S3), and the associated fit parameters are plotted as a function of $p$ in Fig. S2. (b) The mean squared stress fluctuations $\langle \delta\sigma(\gamma, t)^2\rangle_t$ are largest at the $p$- (and thus $z$-) dependent critical strain. The solid lines correspond to the difference between the measured strain-dependent affine and equilibrium differential shear moduli, $K_\mathrm{aff}$ and $K_\mathrm{eq}$, according to Eq. \ref{eqn:stressfluc}. Both the (a) apparent excess zero-shear viscosity $\delta\eta$ and (b) apparent slowest relaxation time $\tau_\mathrm{R}$ also exhibit large peaks at the $p$-dependent critical strain $\gamma_c$.}
\end{figure}

In Fig. \ref{fig_6}a, we show representative $C(\gamma,\tau)$ data for a single sample (the same sample as in Fig. \ref{fig_3}) under applied strains below, near, and above the critical strain $\gamma_c$. We find that stress fluctuations decay slowly when the applied macroscopic strain is near $\gamma_c$, in contrast to a much faster decay for strains below or above the critical regime. We can quantify this strain-dependent change in relaxation dynamics by integrating the shear stress autocorrelation function $C(\gamma,\tau)$ over a sufficiently long range of lag times $\tau$. In practice, the range of lag times for which we can reliably estimate $C(\gamma,\tau)$ is limited by the simulation run time. Assuming $C(\gamma,\tau)$ is known for lag times below a maximum $\tau_\mathrm{max}$, we can estimate the system's slowest relaxation time as
\begin{equation}\label{eqn:relaxationtime}
    \tau_\mathrm{R}(\gamma, \tau_\mathrm{max}) = \frac{\int_0^{\tau_\mathrm{max}}{\tau C(\gamma,\tau)d\tau}}{\int_0^{\tau_\mathrm{max}}{C(\gamma,\tau)d\tau}}
\end{equation}
In the limit $\tau_\mathrm{max}\to\infty$, this converges to the true slowest relaxation time $\tau_\mathrm{R,0}(\gamma)$. In Fig. \ref{fig_6}b, we plot the apparent slowest relaxation time $\tau_\mathrm{R}$ for the same system, with marker colors indicating the strains plotted in Fig. \ref{fig_6}a. Note that the corresponding stiffening curve from Fig. \ref{fig_3} is also shown in the inset. We observe that $\tau_\mathrm{R}$ grows substantially as the critical strain $\gamma_c$ is approached from either side, reaching a maximum at $\gamma_c$. In fact, we see consistent behavior, that is, growth of the slowest relaxation time by an order of magnitude or more with a peak at $p$-dependent critical strain $\gamma_c$, in networks throughout the range of $p$ considered. This is shown in Fig. \ref{fig_7}d, in which the applied strain is normalized by the $p$-dependent critical strain $\gamma_c$, revealing maxima in $\tau_\mathrm{R}$ at $\gamma/\gamma_c=1$ for all $p$.

The stress fluctuation autocorrelation function is also related to the system's lag-dependent excess differential viscosity \cite{wittmer_fluctuation-dissipation_2015} $\delta\eta(\gamma,\tau)\equiv\eta(\gamma,\tau)-\eta_s$ measured at strain $\gamma$,
\begin{equation} \label{eqn:visc}
    \delta\eta(\gamma,\tau) =\int_0^{\tau} C(\gamma,\tau') d\tau',
\end{equation}
which we plot in Fig. \ref{fig_6}c for the same labeled strain values. It is clear that $\delta\eta(\gamma,\tau)$ grows far more dramatically near the critical strain than elsewhere.
In recent work,\cite{shivers_nonaffinity_2022} it was suggested that the low-frequency or ``zero-shear'' excess differential viscosity $\delta\eta_0(\gamma) = \lim_{\tau\to \infty} \delta\eta(\gamma, \tau)$ of prestrained disordered networks is controlled by nonaffinity. Specifically, $\delta\eta_0$ was shown to be related to the quasistatic, athermal differential nonaffinity $\delta\Gamma_\infty(\gamma)$, defined as
\begin{equation}\label{eqn:nonaffinity}
    \delta\Gamma_\infty(\gamma)=\lim_{\delta\gamma\to 0}\frac{1}{N\ell_0^2\delta\gamma^2}\sum_i\left|\delta\mathbf{x}_i^\mathrm{NA}\right|^2
\end{equation}
in which the sum is taken over all network nodes and the vector $\delta\mathbf{x}_i^\mathrm{NA}$ represents the nonaffine component of the displacement vector of node $i$ under a macroscopic, quasistatically applied strain step $\delta\gamma$.  The viscosity-nonaffinity relationship in Ref. \citenum{shivers_nonaffinity_2022} can be stated concisely as
\begin{equation}\label{eqn:nonaff_viscosity}
    \delta\eta_0(\gamma)=\frac{N}{V}\zeta\ell_0^2\delta\Gamma_\infty(\gamma).
\end{equation}
The left-hand side of Eq. \ref{eqn:nonaff_viscosity} reflects the dynamics of the stress fluctuations at finite temperature, while the right side reflects the heterogeneous nature of the strictly quasistatic deformation field, which can be obtained by comparing energy-minimized, $T=0$ system configurations under varying applied $\gamma$.

Simulations have suggested that, for disordered filament networks, the differential nonaffinity generically reaches a maximum at the structure-dependent critical strain,\cite{rens_nonlinear_2014,sharma_strain-driven_2016} at which the system macroscopically transitions between bending-dominated (or floppy, for $\kappa=0$) and stretching-dominated regimes. Thus, according to Eq. \ref{eqn:nonaff_viscosity}, we should generically see a proportional peak in the excess differential viscosity at $\gamma_c$. Testing this prediction requires measuring the quasistatic, athermal differential nonaffinity. To do so, for a given network at strain $\gamma$, we first obtain the energy-minimizing configuration at $T=0$ using the conjugate gradient method. To this configuration, we then apply a small additional strain step $d\gamma = 0.01$, after which we repeat the energy minimization procedure. Comparing the positions in the energy-minimizing configurations at $\gamma$ and $\gamma + d\gamma$, we compute the quasistatic differential nonaffinity $\delta\Gamma_\infty$ using Eq. \ref{eqn:nonaffinity}.  In Fig. \ref{fig_6}d, we plot $\delta\Gamma_\infty$ along with the apparent excess differential zero-shear viscosity for the same system, demonstrating excellent agreement with Eq. \ref{eqn:nonaff_viscosity}.  To demonstrate that this behavior is preserved as the structure is varied, we show in Fig. \ref{fig_7}c that a clear peak in $\delta\eta$ occurs at $\gamma/\gamma_c = 1$ over the entire range of $p$ considered.

We can also use the stress fluctuations to determine the differential relaxation modulus,\cite{wittmer_shear-strain_2015}
\begin{equation}
    K(\gamma,\tau)=K_\mathrm{eq}(\gamma)+C(\gamma,\tau),
\end{equation}
which quantifies the time-dependent apparent stiffness of the system at prestrain $\gamma$ in response to an instantaneous additional strain step.\cite{rubinstein_polymer_2006,doi_theory_1988}
For sufficiently short times, the differential relaxation modulus approaches an upper limit of $\lim_{\tau\to 0}K(\gamma,\tau) = K_\mathrm{aff}(\gamma)$, corresponding to the apparent stiffness of the energy-minimized, athermal equivalent of the system at strain $\gamma$ under an instantaneous, homogeneous infinitesimal shear strain step. Consequently, the equilibrium stiffness of the system under applied strain $\gamma$ is simply $K_\mathrm{aff}(\gamma)$ reduced by $C(\gamma,0)$,\cite{squire_isothermal_1969,lutsko_generalized_1989,wittmer_shear_2013} or equivalently
\begin{equation} \label{eqn:stressfluc}
    K_\mathrm{eq}(\gamma)=K_\mathrm{aff}(\gamma)-\frac{V}{k_B T}\langle\left(\delta\sigma(\gamma,t)\right)^2\rangle_t.
\end{equation}
We find that this relationship provides a useful estimate of the mean squared stress fluctuations, as shown in Fig. \ref{fig_7}b, and we observe that the mean squared stress fluctuations robustly reach a maximum at $\gamma/\gamma_c = 1$.

\section{Discussion and conclusions}

We have investigated the assembly and mechanical testing of disordered networks of crosslinked semiflexible polymers via Brownian dynamics simulations. Such networks serve as essential mechanical constituents of a wide variety of biological materials spanning many length scales. We explored the structural and rheological consequences of varying the crosslinker coverage fraction $p$, an analog of the experimental ratio between the concentrations of crosslinker and filament proteins. Using physical parameters intended to mimic the cytoskeletal polymer F-actin, we measured the effects of increasing the linker coverage fraction $p$ on the average connectivity $z$ of self-assembled networks, observing a corresponding increase in connectivity between $z\in[2.8,3.6]$.

We then investigated the relationship between these connectivity changes and the associated changes in the strain-dependent rheological properties of the assembled networks, obtaining extended time series measurements of fluctuating shear stress $\sigma(\gamma, t)$ for systems held under fixed shear strain $\gamma$. Analyzing many such trajectories gathered over a range of applied strains $\gamma$, we computed the strain-dependent differential shear modulus $K_\mathrm{eq}$, from which we extracted both the linear shear modulus $G_0$ and the critical strain $\gamma_c$. We described the dependence of these quantities on the crosslinker coverage fraction $p$, demonstrating qualitative agreement with the experimentally observed dependence of the same quantities on the crosslinker concentration ratio $R$ reported in Ref. \citenum{gardel_elastic_2004}. Specifically, we found that increasing $p$ produces inherently stiffer networks (having an increased linear shear modulus $G_0$) that simultaneously exhibit an earlier tendency to strain-stiffen (having a lower critical strain $\gamma_c$). Drawing comparisons between the scaling of both $G_0$ and $\gamma_c$ with $p$ and the analogous experimentally observed scaling of the same quantities with $R$, we suggested a simple power law relationship between $p$ and $R$. We then described the apparent dependence of these elastic features on the $p$-dependent structural quantities $z$ and $\ell_c$, notably showing that the critical strain decreases linearly with $z$. A linear fit suggests that $\gamma_c$ approaches $0$ as the connectivity $z$ approaches a limiting value $z^*\to z_c$ near $4$, the theoretical upper bound for $z$ at high crosslinking density.
 
We then extended our observations beyond strictly elastic properties by analyzing the stress fluctuation autocorrelation function $C(\gamma,\tau)$, which revealed the development of extremely slow dynamics in systems subjected to applied shear strains near the critical strain $\gamma_c$.  From $C(\gamma,\tau)$, we obtained estimates of the slowest relaxation time $\tau_{\mathrm{R},0}$ and the excess differential zero-shear viscosity $\delta\eta_0$, both of which we showed are consistently maximized at the $p$-dependent critical strain, irrespective of the details of the underlying network. Building upon results from Ref. \citenum{shivers_nonaffinity_2022}, we demonstrated that the excess differential viscosity in these finite-temperature systems is quantitatively controlled by the athermal quasistatic differential nonaffinity $\delta\Gamma_\infty$, a measure of the inherent tendency of the strained network to deform heterogeneously.  Importantly, our results suggest that one should expect measurable dynamical signatures of transition-associated nonaffine fluctuations to appear in semiflexible polymer networks with biologically relevant elastic properties, e.g. those of the F-actin cytoskeleton, under physiologically relevant applied strains. In other words, one should generically expect to observe slowing stress relaxation in biopolymer networks at prestrain levels near the critical strain $\gamma_c$, marking the macroscopic transition between bending-dominated and stretching-dominated elasticity. Since $\gamma_c$ is controlled by the average connectivity $z$, which is controlled in turn by the availability of crosslinking sites (here, $p$), our results suggest a route to experimentally control the strain dependence of major features of both network elasticity and stress relaxation dynamics.

 In future work, it would be prudent to explore how the behavior we observe near the critical strain might differ in networks with linkers capable of either breaking under sufficient tension \cite{burla_connectivity_2020,lwin_rigidity_2022,tauber_stretchy_2022} or  transiently binding and unbinding.\cite{amuasi_linear_2018,scheff_actin_2021} In such systems, we anticipate a rich interplay between the slow relaxations associated with nonaffine rearrangements and the additional dynamics of network remodeling and fracture.

\section*{Supporting Information}
Simulation parameters; connectivity and inter-crosslink contour length calculation details; discussion of fitting the differential shear modulus; discussion of the stopping criterion for network assembly; sample animation of network assembly

\begin{acknowledgments}

This study was supported in part by the National Science Foundation Division of Materials Research (Grant No. DMR-1826623) and the National Science Foundation Center for Theoretical Biological Physics (Grant No. PHY-2019745). This work began as a summer research experience for undergraduates (REU) project supported by the Frontiers in Science (FIS) program of the Center for Theoretical Biological Physics.

\end{acknowledgments}

\bibliographystyle{achemso}
\bibliography{Bibliography}

\end{document}

\title[]{Supporting Information -- Structural features and nonlinear rheology of self-assembled networks of cross-linked semiflexible polymers}
\author{Saamiya Syed}
\affiliation{ 
College of Technology, University of Houston, Houston, TX 77204, USA
}%
\affiliation{ 
Center for Theoretical Biological Physics, Rice University, Houston, TX 77005, USA
}%
\author{Fred C. MacKintosh}%
\affiliation{ 
Center for Theoretical Biological Physics, Rice University, Houston, TX 77005, USA
}%
\affiliation{ 
Department of Chemical and Biomolecular Engineering, Rice University, Houston, TX 77005, USA
}%
\affiliation{ 
Department of Chemistry, Rice University, Houston, TX 77005, USA
}%
\affiliation{ 
Department of Physics \& Astronomy, Rice University, Houston, TX 77005, USA
}%
\author{Jordan L. Shivers}
\email[]{jshivers@uchicago.edu}
\affiliation{ 
Center for Theoretical Biological Physics, Rice University, Houston, TX 77005, USA
}%
\affiliation{ 
Department of Chemical and Biomolecular Engineering, Rice University, Houston, TX 77005, USA
}%
\maketitle
\thispagestyle{plain}

\onecolumngrid

\FloatBarrier
\section{Calculation of the connectivity and average contour length}\label{appendix_connectivity}

\begin{figure}[htb!]
\centering
\includegraphics[width=0.5\columnwidth]{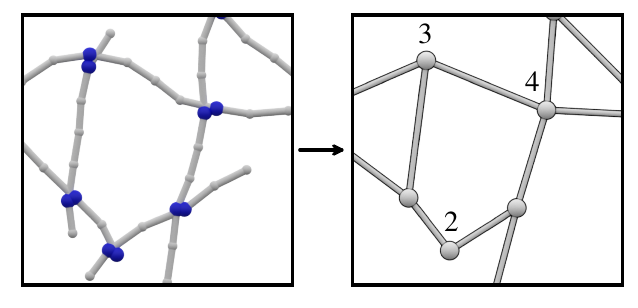}
\caption{ \label{figS1} To calculate a network's average connectivity $z$, we consider a reduced network structure that treats each crosslink as a node and each inter-crosslink (non-dangling) filament section as an edge. On the right, a few example nodes in the reduced network are labeled according to their connectivity. 
}
\end{figure}

To estimate the connectivity of a given assembled structure, we consider a reduced network defined by (1) merging each pair of crosslinked “linker” nodes into a single vertex and then (2) merging each filament section separating two vertices into a single edge. This is sketched in Fig. \ref{figS1}. 

\FloatBarrier
\section{Parameters}\label{appendix_parameters}

In the interest of drawing comparison to experimental results in Ref. \citenum{gardel_elastic_2004}, we have chosen parameter values roughly appropriate for F-actin in laboratory conditions. Specific values are listed in Table \ref{table1}.

\begin{table}[htb!]
\begin{tabular}{|c|c|c|}
\hline
\textbf{quantity}                  & \textbf{symbol}                  & \textbf{value} \\ 
\hline
number of filaments                & $N_f$                         & $500$ \\
filament length       & $\ell_f$                         & $9\;\si{\micro\meter}$  \\
filament length per   volume       & $\rho$                           & $2.6\;\si{\micro\meter}^{-2}$ \\
number of nodes per filament       & $n$                              & $10$ \\

filament persistence length                 & $\ell_p$                         & $17\;\si{\micro\meter}$ \\
crosslink rest length              & $\ell_{0,c\ell}$                     & $0.2\;\si{\micro\meter}$ \\
filament stretching   rigidity     & $\mu$                            & $588\;\si{\pico\newton}$   \\
thermal energy scale               & $k_B T$                          &  $4.11\times 10^{-3}\;\si{\pico\newton}\cdot\si{\micro\meter}$      \\
solvent viscosity                  & $\eta_s$                         & $0.001\;\si{\pico\newton}\cdot\si{\micro\meter}^{-2}\cdot \si{s}$\\
timestep                           & $\Delta t$                       &  $9.42\times10^{-7}\;\mathrm{s}$ \\
total time, assembly       & $\tau_\mathrm{a}$                         & $60\;\mathrm{s}$   \\ 
total time, rheology      & $\tau_\mathrm{tot}$                     & $30\;\mathrm{s}$ \\    
\hline
\end{tabular}
\caption{ \label{table1} Independent parameters, in real units.}
\end{table}

\begin{table}[htb!]
\begin{tabular}{|c|c|c|}
\hline
\textbf{quantity}                  & \textbf{symbol}                  & \textbf{value} \\ 
\hline
filament bond rest length          & $\ell_0=\ell_f/(n-1)$                         & $1\;\si{\micro\meter}$ \\
simulation box edge   length       & $L=(N_f\ell_f/\rho)^{1/3}$                              & $12\;\si{\micro\meter}$ \\
filament bending rigidity & $\kappa = k_B T \ell_p$ & $6.99\times10^{-2}\;\si{\pico\newton}\cdot\si{\micro\meter}^2$ \\
\hline
\end{tabular}
\caption{ \label{table2} Parameter-dependent quantities.}
\end{table}
 Assuming a molecular mass of $42 \;\si{\kilo\dalton}$ for actin\cite{ballweber_polymerisation_2002}, the mass per length is $\sim 16\times10^3\; \si{\kilo\dalton}/\si{\micro\meter}$, which translates to $0.63\;\si{\micro \mol}/\si{\micro\meter}$. Therefore, our chosen length density of $\rho=2.6\;\si{\micro\meter}^{-2}$ corresponds to an actin concentration of $c_A = 1.63\;\si{\micro M}$.

Quantities used in simulations are nondimensionalized by characteristic length, force, and drag coefficient values: $\ell^*=1\;\si{\micro\meter}$, $f^*=1\;\si{\pico\newton}$, and $\zeta^*=6\pi\eta_s(\ell_0/2)=9.42\times10^{-3}\;\si{\pico\newton}\cdot\si{\micro\meter}\cdot \mathrm{s} $ (for example, the simulation timestep is $\Delta t_\mathrm{sim}=\Delta t/(\zeta^* \ell^*/f^*)=10^{-4}$.  Note that one can define a dimensionless bending rigidity $\tilde{\kappa}=\kappa/(\mu\ell_0^2)$, as in past work.\cite{sharma_strain-controlled_2016} With our parameters, $\tilde{\kappa}=1.18\times10^{-4}$.

\section{Fits of K to the expected scaling form}\label{appendix_fits}
We assume that the differential shear modulus $K=\partial\sigma/\partial\gamma$ can be written as
\begin{equation} \label{Eq1}
K = a\mathcal{K}(\tilde{\kappa},\gamma)
\end{equation}in which $a$ is some prefactor with units of stress, $\tilde{\kappa}$ is a dimensionless bending rigidity, and $\mathcal{K}$ is a dimensionless function of strain that scales with the distance to a critical strain $\gamma_c$ as\cite{sharma_strain-controlled_2016}
\begin{equation}
\mathcal{K} \propto|\gamma-\gamma_c|^f\mathcal{G}_\pm\left(\frac{\tilde{\kappa}}{|\gamma-\gamma_c|^\phi}\right).
\end{equation}This scaling is reproduced by the constitutive equation
\begin{equation}
\frac{\tilde{\kappa}}{|\gamma-\gamma_c|^\phi} \sim \frac{\mathcal{K}}{|\gamma-\gamma_c|^f}\left(\pm1 + \frac{\mathcal{K}^{1/f}}{|\gamma-\gamma_c|}\right)^{\phi-f}
\end{equation}
in which the plus and minus correspond to the regions below and above the critical strain, respectively.
We compute $\mathcal{K}(\gamma)$ by numerical inversion of $\tilde{\kappa}\left(\mathcal{K}\right)$:
\begin{equation}\label{eq4} \tilde{\kappa}=b\mathcal{K}\left(\pm|\gamma-\gamma_c|+\mathcal{K}^{1/f}\right)^{\phi-f},
\end{equation}
setting $b=1$. 

\begin{figure}[htb!]
\centering
\includegraphics[width=0.5\columnwidth]{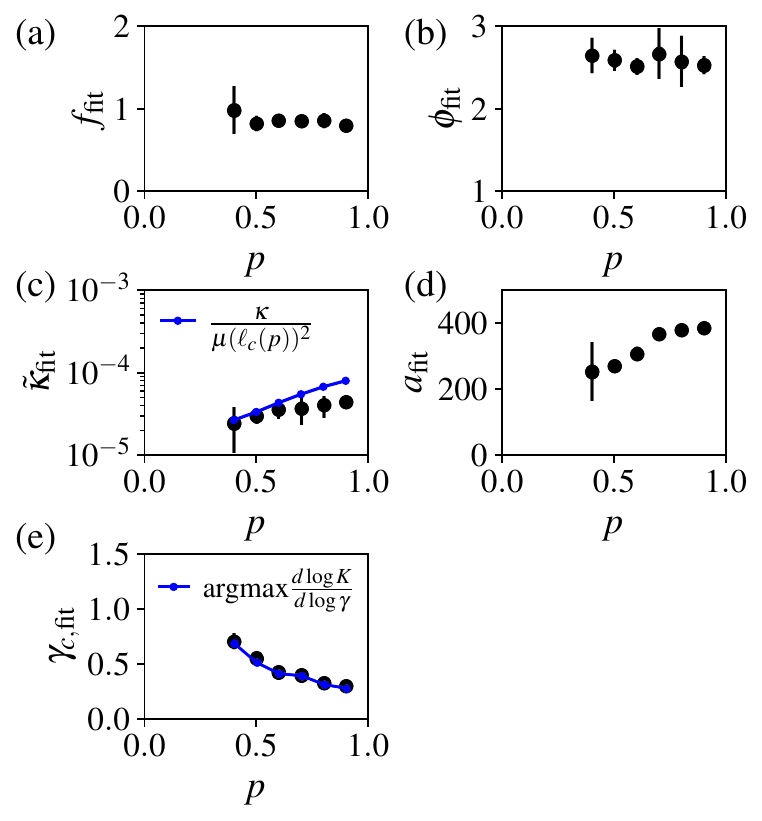}
\caption{ \label{figS2} Best-fit parameters for systems with varying crosslinker coverage fraction $p$, corresponding to the solid curves in Fig. 7a in the main text.}
\end{figure}

For each set of measured $K(\gamma)$, we determine the best-fit parameters $\{\tilde{\kappa},\gamma_c,f,\phi,a\}$ by performing a fit of $\log_{10}(a\mathcal{K}(\gamma))$ to $\log_{10}(K(\gamma))$ using the non-linear least squares fitting function \texttt{scipy.optimize.curve\_fit} from the open source Python library SciPy \cite{virtanen_scipy_2020}.  Best-fit parameters for the data in Fig. 7a in the main text are shown in Fig. \ref{figS2}. We note that one can alternatively compute the critical strain $\gamma_c$ as the value of $\gamma$ corresponding to the inflection point of $\log_{10}K$ vs. $\log_{10}\gamma$,
\begin{equation}\label{gammac}
\gamma_c = \mathrm{argmax}\frac{d\log_{10}K}{d\log_{10}\gamma}
\end{equation} as has been done in related work \cite{sharma_strain-controlled_2016}. In Fig. \ref{figS2}e, we show that the values of $\gamma_c$ determined using the fitting procedure described here agree well with values computed using Eq. \ref{gammac}.

\FloatBarrier
\section{Stopping criterion for network assembly}

\begin{figure}[htb!]
\centering
\includegraphics[width=0.6\columnwidth]{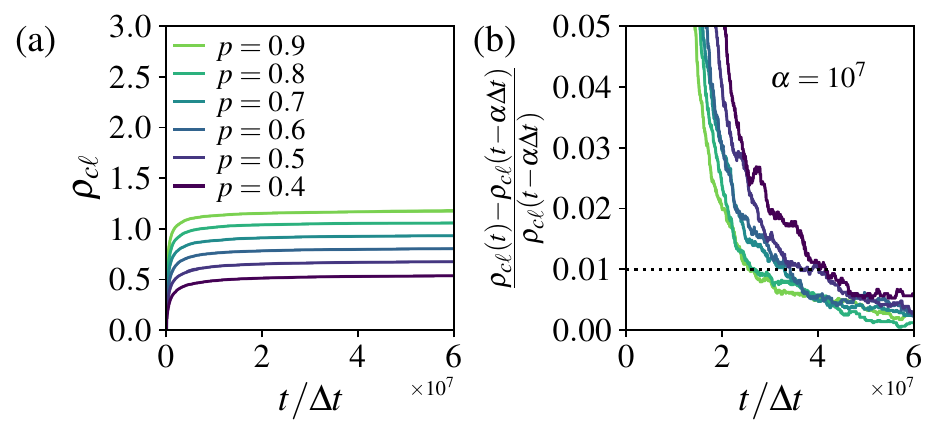}
\caption{ \label{figS3} (a) Total number of crosslinks formed per volume, $\rho_{c\ell}$, as a function of elapsed assembly time. (b) Fractional increase in the total number of crosslinks formed per volume between times $t-\alpha\Delta t$ and $t$, with $\alpha=10^7 $. The dotted line indicates our network stabilization criterion, $(\rho_{c\ell}(t) - \rho_{c\ell}(t-\alpha\Delta t))/\rho_{c\ell}(t-\alpha\Delta t)\le 10^{-2}$.}
\end{figure}

The rate at which crosslinks are formed slows as the assembly process progresses. To determine an appropriate stopping time for the assembly process, we track the total number of crosslinks formed per volume, $\rho_{c\ell}$, as a function of elapsed assembly time $t$, as shown in Fig. \ref{figS3}a. We deem network structures "stabilized" if, at time $t$, the total number of crosslinks formed per volume has increased by less than $1\%$ over the most recent $\alpha=10^7$ timesteps. Equivalently, we require that $(\rho_{c\ell}(t) - \rho_{c\ell}(t-\alpha\Delta t))/\rho_{c\ell}(t-\alpha\Delta t) \le 10^{-2}$. In Fig. \ref{figS3}b, we plot the fractional increase, $(\rho_{c\ell}(t) - \rho_{c\ell}(t-\alpha\Delta t))/\rho_{c\ell}(t-\alpha\Delta t)$, as a function of elapsed time. While it is clear that our network stabilization criterion is satisfied by $t/\Delta t \approx 4\times 10^7$ for the chosen range of parameters, we conservatively select a total assembly time $\tau_\mathrm{a} = 6\times 10^7\Delta t$ for all systems.

\bibliographystyle{achemso}
\bibliography{Bibliography}